\documentclass[prl,amsmath,showpacs, amsfonts,twocolumn,preprintnumbers, superscriptaddress]{revtex4} 
\usepackage{amsfonts,amsmath}
\usepackage{color,graphicx} 
\usepackage{units} 
\usepackage{setspace}
\usepackage{bm}
\usepackage{latexsym, amsfonts, amssymb}
\usepackage{fancyhdr}
\newcommand{\etnhfour}{$\alpha$-(ET)$_2$NH$_4$(SCN)$_4$}

\newcommand{\ET}{ET-NH4}
\newcommand{\micro}{\mu }
\newcommand{\ohm}{\Omega }
\newcommand{\hctwo}{H_{c_2}}
\newcommand{\hp}{H_\text{P}}
\newcommand{\hpBCS}{H_\text{P}^{\rm BCS}}
\newcommand{\tc}{T_\text{c}}

\begin{document}
\title {A bulk 2D Pauli-Limited Superconductor}
\author{T. Coffey}
\affiliation{Department of Physics, Clark University, Worcester, MA,  01610}
\author{C. Martin}
\author{C. C. Agosta}
\affiliation{Department of Physics, Clark University, Worcester, MA,  01610}
\author{T. Kinoshota}
\affiliation{CREST, JST, Kawaguchi 332-0012 Japan} 
\author{M. Tokumoto}
\affiliation{CREST, JST, Kawaguchi 332-0012 Japan} 
\affiliation{Nanotechnology Research Institute, AIST, Tsukuba 305-8568, Japan}
\date{\today}

%\begin{document}
\begin{abstract}
We present a nearly perfect Pauli-limited critical
field phase diagram for the anisotropic organic superconductor 
\etnhfour \ when the applied magnetic field is oriented parallel to the conducting layers. The critical fields ($\hctwo$) were found by use of penetration depth measurements. Because $\hctwo$ is Pauli-limited, the size of the superconducting energy gap can be calculated.
The role of spin-orbit scattering and many-body effects play a role 
in explaining our measurements. 

\end{abstract}


\maketitle
%Intro
Over the past 40 years there have been many theoretical calculations \cite{lawrenceDoniach,klemmLB,bulaevskiiJETP73,schneider,lebedPRL}
 and experiments on  layered superconductors \cite{adamsPRL2004,strunkPRB,zuo,manskyPRL93,leePRL,kwok,proberSB}
 subjected to an external magnetic field  applied parallel to the conducting layers.
When the magnetic field is precisely aligned parallel to the conducting layers, magnetic flux lines effectively penetrate the least conducting layers between the conducting planes and  
the orbital destruction of superconductivity associated with the vortices is suppressed.

In this case, the superconductor can be described as a series of Josephson coupled layers~(JCL) as long as the temperature is below the 2-D/3-D transition temperature $(T < T^*)$ that occurs when 
the coherence length perpendicular to the layers($\xi_o^\perp$) is less than the distance between the layers. In the parallel orientation, the Cooper pairs will be broken for a typical superconductor when the
energy gained by being in the superconducting state equals the energy cost of maintaining 
anti-parallel spins in an applied field.
This limit is known as the Clogston-Chandrasekar or Pauli paramagnetic limit. 
The upper critical field ($\hctwo$) for a Pauli-limited superconductor is
$\hp=\frac{\sqrt{2}\Delta_o}{g\mu_\text{B}}$ where 
$\Delta_o$ is the energy gap, $g$ is the Land$\acute{\text{e}}$ g-factor,
and $\mu_\text{B}$ is the Bohr magneton \cite{clogston}.
For a BCS superconductor $\mu_o\hpBCS=\unitfrac[1.84]{T}{K}~\tc$. 
Many investigations of layered superconductors have focused on the possibility of realizing 
an exotic, inhomogeneous state known as the Fulde-Ferrell-Larkin-Ovchinikov state 
(FFLO) \cite{fuldeFerrell,larkinOvchinikuv}, which allows $\hctwo$ to exceed $\hp$.

% State the point
In this letter we present a Pauli-limited critical field phase
diagram with the applied field parallel to the conducting layers  ($\mu_oH\parallel ac$-plane)
 for the  anisotropic organic superconductor \etnhfour (short form \ET) ~\cite{motivation}. 
Our phase diagram  is based on rf penetration depth measurements that utilize a tunnel diode
 oscillator (TDO) \cite{coffeyRSI, mielkeSingleton}.  The phase diagram we present is a close match to the calculations from Klemm, Luther, and Beasley \cite{klemmLB} in the limit of small spin-orbit scattering, and the low-temperature $\hctwo$ we measure is a close match to the many-body prediction from McKenzie \cite{mckenzie}. The remarkable result is that due to Pauli-limiting, $\hctwo$ tracks the superconducting energy gap at all temperatures.  In contrast to other organic superconductors\cite{AgostaMartin2006,ConiglioWinter2010PRB, ChoSmith2009PRB, LortzWang2007PRL}, we observe no evidence of $\hctwo$ exceeding the Pauli limit in \ET.

%Segway into explaining why NH4 is important
Previous attempts to synthesize and measure a bulk Pauli-limited superconductor
have been stymied by materials that have low anisotropy or weak spin-orbit scattering \cite{strunkPRB}.
Other investigations may have been limited by the available magnetic fields and low temperatures at the 
time of the experiments \cite{proberSB}. A convincing Pauli-limited superconductor was realized for a single layer of aluminum \cite{tedrowPRB73}.

Despite having a low $\tc$ ($\approx\unit[0.9]{K}$  \cite{andraka,nakazawa,taniguchi1,wang}), 
\ET\ is an ideal material to study because it is one of the most anisotropic layered superconductors
and should be a model JCL superconductor. The anisotropy parameter ($\gamma$) for \ET~=~2000, while $\gamma$=150 for Bi$_2$Sr$_2$CaCu$_2$O$_{8+\delta}$ \cite{taniguchi1}.  The high anisotropy in \ET\ produces a broad, zero-field resistive transition
(reported  onset temperatures vary from $\unit[1.2]{K}$ 
to $\unit[2.4]{K}$ \cite{brooks,shimojoJPSJ2002,wang,shimojo99,sato,taniguchi1}),
a broad peak in the zero-field specific heat at $\tc$ \cite{andraka,nakazawa,taniguchi1},
anisotropic penetration depths  ($\lambda_\perp=\unit[1400]{\micro m}$ and $\lambda_\parallel=\unit[0.7]{\micro m}$ as $T \rightarrow 0$ \cite{taniguchi1}, and anisotropic resistivity ($\unit[20]{\ohm\cdot cm}\leq \rho_\perp \leq  \unit[40]{\ohm\cdot cm}$ and $\unit[10]{\micro\ohm\cdot cm} \leq \rho_\parallel\leq\unit[100]{\micro\ohm\cdot cm}$ at $\unit[3]{K}$ \cite{taniguchi1}). 

%end Importance of NH4

%begin  Global Experimental details
In a TDO experiment we measure the  amplitude ($A$) and frequency shift ($\Delta F$)
of a self-resonant circuit with unperturbed frequency $F_o$. These quantities are determined  by the complex impedance of the coil containing the sample ($L^{\prime}$). For the case of a long rod in a axial coil, $L^{\prime}=L_0\left(1+4\pi\eta\{\chi^{\prime}-\text{j}\chi^{\prime\prime}\}\right)$,
where $\eta$ is defined as the volume filling factor, directly related to the penetration depth. As the rf field is expelled from the sample, $\eta$ changes and thus $A$ and $\Delta F$, related to $\chi^{\prime}$ and $\chi^{\prime\prime}$ shift together.   The region where 
$\eta$ dominates is called the skin depth region \cite{gruner1}.   The sample we characterized is a small block 
($\unit[1.69]{mm}$ by $\unit[1.83]{mm}$ by $\unit[1.0]{mm}$), not a long rod;
therefore, $L^{\prime}$ must be corrected by a demagnetization factor in order to measure
absolute quantities. 
 We did not measure the demagnetization factor of our sample and do not report absolute 
values for the penetration depth. However, the temperature or applied field at which a transition occurs
is clear in our measurements.

In type II superconductors, a complex penetration depth ($\widetilde{\lambda}$)  includes the normal skin depth ($\delta$), the London penetration depth ($\lambda_L$), 
and the Campbell penetration depth ($\lambda_c$) \cite{coffeyClemPRL91,sridhar,manskyPRL93}.   
As the applied field is increased the following occurs: the flux lattice is destroyed, $\lambda_L$ diverges
at $\hctwo$, and $\tilde{\lambda}$ becomes limited by $\delta$.
From our data, we define $\hctwo$ via the maximum in the second derivative 
of the frequency response of the TDO ($\Delta  F$) with respect to the applied field and
is calculated for fields above and below $\hctwo$.
%end  Global Experimental details
 
\begin{figure} \begin{center}
	\includegraphics[keepaspectratio=1,width=8 cm]{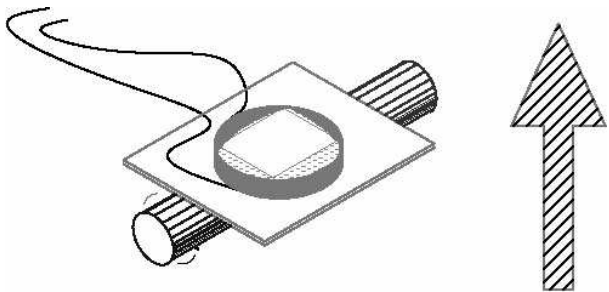} 
	\caption{\label{fig:platform} Schematic of coil and sample on a rotating platform.
	The large arrow indicates the direction of the applied field.
	The cylinder is the axis of rotation. }
	\vspace{-0.7cm}
\end{center} \end{figure}

%begin particular experimental details
The phase diagram we present is constructed from three different experimental runs, 
two at the NHMFL and one at Clark University,
using three different circuits, where $F_o\approx\unit[25]{MHz},\unit[10]{MHz},\unit[35]{MHz}$ respectively. All three experimental runs produced consistent results.
The parallel orientation is determined from the sharp cusp in $\hctwo$ verses angle.
At Clark University, a transverse $\unit[1]{T}$ electromagnet,
single-shot He$^3$ refrigerator, and rotating probe with an angular 
resolution of $\unit[0.1]{^\circ}$ were used. 
At the NHMFL a $\unit[18]{T}$ superconducting magnet, dilution refrigerator, and 
rotating probe with an  angular resolution of $\unit[0.050]{^\circ}$ were used \cite{murphyRot}.
In all cases the sample is placed in a small coil that is part of a self-resonant circuit. 
Small balls of cotton are packed between the sample and the coil and a Teflon
gurney is tied  around the coil and the rotating platform to hold
the sample in place.  The flattest side of the sample sits on a rotating platform
such that the conducting ($ac$) planes are roughly parallel to the platform and perpendicular to the axis of the coil, as in Fig.~\ref{fig:platform}. The platform's axis of rotation is roughly parallel to
either the $a$ or $c$ crystallographic direction of the sample.
The rf coil excites currents in the conducting planes of the sample. 
Using $\rho_\parallel$ from Ref.~\cite{taniguchi1},
$\delta_\parallel=\unit[70]{\micro m}$ at $\unit[25]{MHz}$ and $\unit[3]{K}$.
The rf field generated by the coil is $\lesssim\unit[70]{\micro T}$. 
%Fig 1

%end Experimental details

%begin Data
In addition to the parallel and perpendicular phase diagrams, we found $\tc\approx\unit[0.96]{K}$
from temperature sweeps at Clark University.  In the normal state the TDO measures resistivity, which as a function of magnetic field exhibits Shubinkov-de Haas (SdH) oscillations \cite{shoenburg}. From these SdH oscillations we extracted Fermi surface parameters such as: 
the SdH frequency, $F_\text{SdH}=\unit[564\pm2]{T}$, the effective mass, $m^*=2.5 m_e$, and  the Dingle temperature $T_\text{D}=\unit[1.11]{K}$, which can be directly translated into a scattering rate via $\frac{1}{\tau}=2\pi k_b T_D/\hbar$.   From our SdH data, we find $\ell=\unit[681]{\AA}$ and $\tau=\unit[1.09]{ps}$, where $\ell$ is the mean free path of the conducting quasi-particles.  These results are consistent with previous results \cite{wosnitzaPRB}.
From the perpendicular, zero-temperature critical field ($\mu_o\hctwo^\perp(0)=\unit[0.13]{T}$) we calculated the coherence length for quasi-particles in the conducting layers, $\xi_o^\parallel=\unit[500]{\AA}$.  To the best of our knowledge, a previously reported value ($\xi_o^\parallel=\unit[500]{\AA}$) 
was calculated from the slope ($\frac{d(\mu_o\hctwo)}{dT}|_{\tc}=\unitfrac[-0.08\pm 0.02]{T}{K}$) 
of a perpendicular phase diagram generated via specific heat data \cite{taniguchi1}.
For a superconductor, the dirty limit is defined when $\ell<<\xi_o^\parallel$. 
Given that  $\ell/\xi_o^\parallel \approx 1.4$, \ET\ resides just on the clean side of the 
boundary between a clean and dirty superconductor.
 
\begin{figure}[!] \begin{center}
	\includegraphics[keepaspectratio=1,width=6 cm, angle = 90]{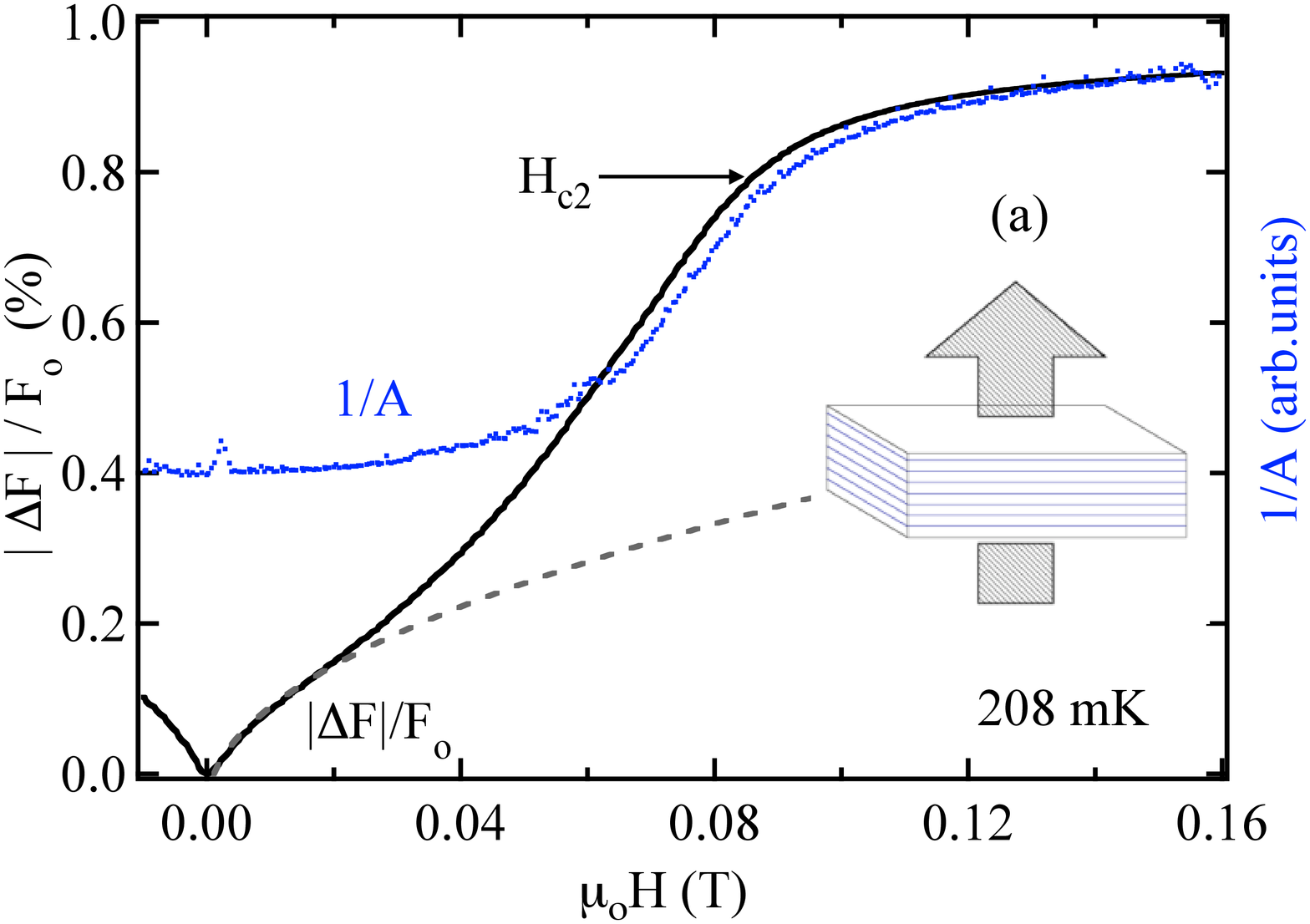} 
	\includegraphics[keepaspectratio=1,width=6 cm, angle = 90]{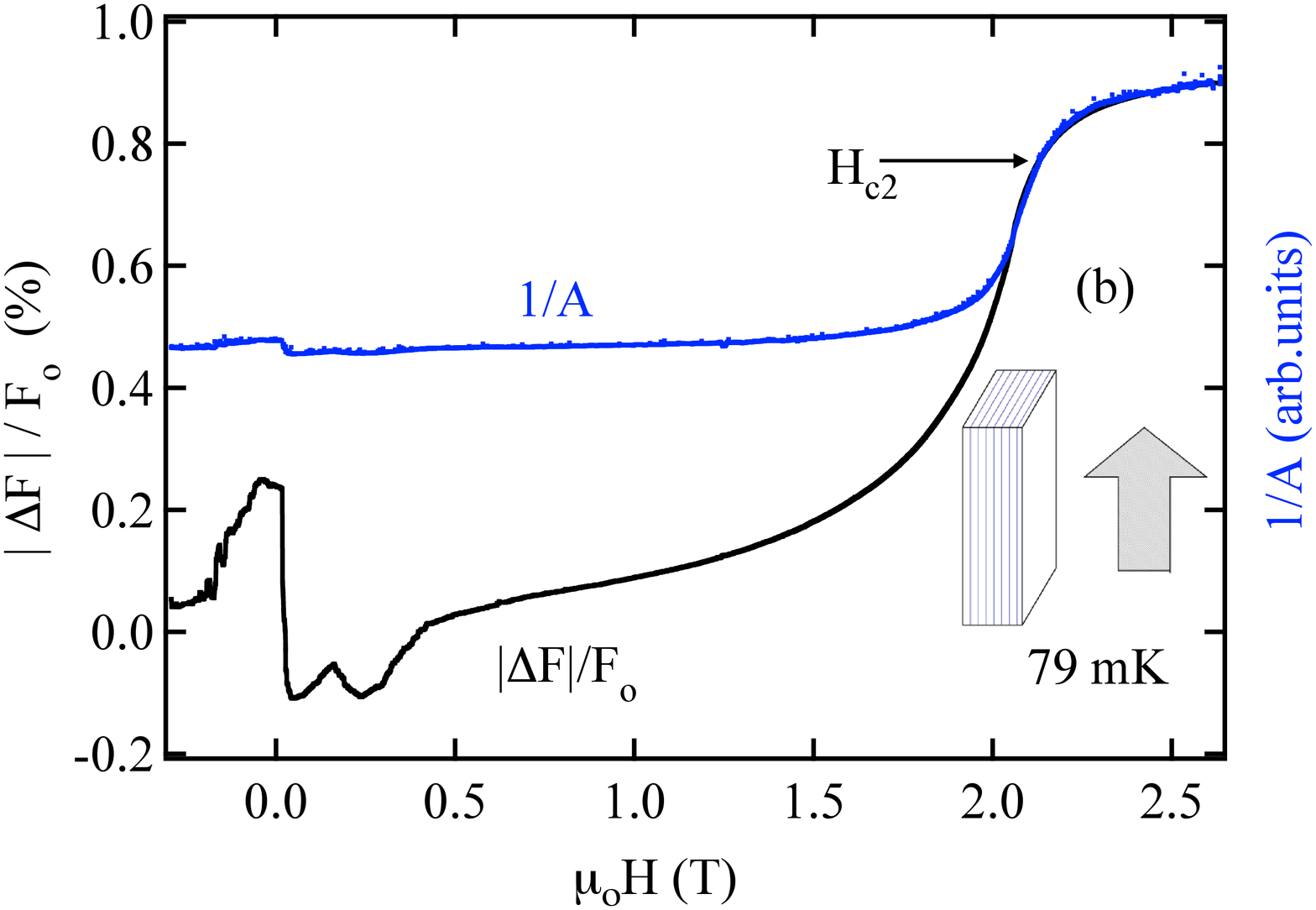} 
	\caption{\label{fig:fandAPerp} Comparison of $\frac{|\Delta F|}{F_0}$ and $\frac{1}{A}$ versus applied field.  As a result of being in the skin depth limit, just below and above $\hctwo$, $\frac{1}{A}$, and $\frac{|\Delta F|}{F_0}$ change in an almost identical manner.
	The dashed line in Fig. 2a is a square root fit to the low field frequency data. 
	In order to emphasize the correlation between the frequency and amplitude data $|\Delta F|$ is plotted
	instead of $\Delta F$. As the field increases the frequency actually decreases. 
	 The insert shows the orientation
of the applied field with respect to the conducting layers. 
}
\end{center} 
\vspace{-0.7cm}
\end{figure}

Fig.~\ref{fig:fandAPerp} shows amplitude and frequency data for field sweeps in the parallel and perpendicular orientations.  This experiment is in the skin depth regime because  $\delta/r_s \approx 0.04$.  Fig.~\ref{fig:fandAPerp}a suggests that we observe Campbell penetration in the perpendicular orientation because $\frac{\Delta F}{F_0} \propto \sqrt H$ at low fields.
There is very little structure in the amplitude signal at low field, because the coil resistance dominates $Q_\text{tc}$. 

%Fig 2 A and B
%end Data

%begin Theoretical wrapper and discussion of our data
Microscopic calculations in the clean \cite{bulaevskii}
and dirty \cite{klemmLB} limits have predicted specific phase diagrams for JCL superconductors.
For a clean, JCL superconductor with negligible spin-orbit scattering ($\tau_{so}\rightarrow\infty$),
Ref.~\cite{bulaevskii} predicts that $\hctwo^\parallel=\hpBCS\sqrt{1-T/\tc}$ when  
the temperature is below the 2-D/3-D transition ($T<T^*)$.
For a dirty, JCL superconductor with strong spin-orbit scattering ($\tau_\text{so}\rightarrow 0$),
$\hctwo$ may reach up to 6 times $\hp$ \cite{klemmLB} according to 
\begin{equation}
\hctwo^\parallel(0) = 0.602(k_b\tc\tau_{\rm so}/\hbar)^{-1/2}\hpBCS.
\label{eq:spinOrbit}
\end{equation}
In a dirty system with no spin-orbit scattering($\tau_{so}\rightarrow\infty$),
$\hctwo$ saturates near $\hp$ \cite{klemmLB}. 

A relevant complication is that two mechanisms can enhance $\hpBCS$ --- strong coupling and 
many-body effects. Many-body effects can increase $\hp$ for a spin-singlet Fermi liquid with a quasi 2-D circular Fermi surface by the factor $R$, $\hp=\frac{\hpBCS}{R}$, where $R$ is Wilson's ratio \cite{mckenzie}. A theory-independent method to estimate the many-body enhancement has also been proposed \cite{zuo, agostaPauli}.   This method equates the condensation energy of the superconductor and the energy gained by having the electrons' spin align with the applied field. 

The low concentration of conducting quasi-particles in the organic superconductors promotes strong interactions among the carriers and $\hp$ may well be enhanced by many-body effects. 
Wilson's ratio ($R$) for \ET\ is $0.7\pm0.2$ or $0.86\pm 0.05$ \cite{mckenzie}.   
Using the value of $R$ with the smaller uncertainty and $\hpBCS = 1.77$, $\frac{\hp}{\hpBCS}=1.17\pm0.07$.
Using the jump in specific heat data \cite{andraka,nakazawa,taniguchi1} 
to estimate the condensation energy and susceptibility \cite{miyagawa}, 
we estimate $\hp$ using the theory independent method \cite{zuo}.
This method depends on which set of specific heat data is used, and produces a high
and low estimate very similar to the many-body theory using Wilson's ratio. 
In both cases the low estimates predict $\frac{\hp}{\hpBCS}=1.2$, which 
matches the ratio of the measured zero-temperature $\hctwo$ and $\hpBCS$.

The inherent assumptions in the JCL theories presented in Ref.~\cite{klemmLB} and Ref.~\cite{bulaevskii}
assume a weakly coupled, $s$-wave superconductor.
In contrast to many other organic superconductors, 
specific heat data show that \ET\ is a weakly coupled superconductor \cite{wankaPRB98}. 
Even though $d$-wave symmetry has been proposed in the 
$\kappa$ phase organic superconductors and $p$-wave symmetry clearly shown  in some Bechgaard salts \cite{leePRL}, 
there has been no data indicating anything besides a conventional order parameter in \ET. 

The inherent 2D nature of our sample suggests that it is below $T^*$ over most of the temperature range. It is expected that 
the angular dependence of $\hctwo$ follows 
the bell-like anisotropic G-L equation,
$\hctwo^2\left[\left(\frac{\cos(\Phi)}{{\hctwo}_\perp}\right)^2 +	
	\left(\frac{\sin(\Phi)}{{\hctwo}_\parallel}\right)^2\right]=1$,
above $T^*$ and 
the cusp-like Tinkham thin film equation,
\begin{equation}
	\left|\frac{\hctwo(\Phi)\cos{(\Phi)}}{{\hctwo}_\perp}\right|+
	\left[\frac{\hctwo(\Phi)\sin{(\Phi)}}{{\hctwo}_\parallel}\right]^2=1,
			\label{eq:g3Theory}
\end{equation}
below $T^*$\cite{schneider}.

\begin{figure}[!] \begin{center}
	\includegraphics[keepaspectratio=1,width=6 cm, angle = 90]{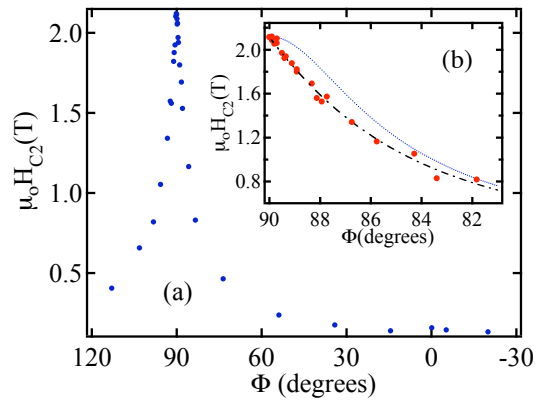} 
	\vspace{-0.2cm}
	\caption{\label{fig:hc2Angle}  $\hctwo$ as a function of angle at $\unit[40]{mK}$.
          Inset: The dot-dashed line
            is a fit to Eq.~(\ref{eq:g3Theory}) and the solid line is a fit to the 
	anisotropic G-L theory.  The angle($\Phi$) between the normal of the 
	conducting layer and the applied field is zero when the applied field is 
	perpendicular to the conducting layers.}
	\vspace{-0.7cm}
	\end{center} \end{figure}

For a truly Pauli limited superconductor even Eq. 2 should not be valid, because it is still based on orbital destruction of superconductivity. However,  one can argue phenomenologically that a similar equation should exist with H$_P$ in place of $\hctwo$ parallel. A theoretical argument for this phenomenological equation was made by Bulaevskii.\cite{Bulaevskii90JMPB} When the sample is within a fraction of a degree of parallel, the $\hctwo$ should deviate form any of these equations, and this phenomena was seen in recent studies \cite{ConiglioWinter2010PRB}. 
To insure that the parallel critical fields we report are accurate, seven
full angular studies were conducted between $\unit[40]{mK}$ and  $\unit[750]{mK}$. 
Below $\unit[750]{mK}$ our data fit very well to the Tinkham thin film equation (Eq.\ref{eq:g3Theory}).
Fig.~\ref{fig:hc2Angle}a plots $\hctwo$ as a function of angle at $\unit[40]{mK}$.
A comparison between the data and the predictions
for a layered superconductor in the 2D and 3D limiting cases is made in
Fig.~\ref{fig:hc2Angle}b at angles close to the parallel orientation.
The agreement between our data and the Tinkham thin-film equation
indicates that the vortices in each layer are decoupled and 
\ET\ is a JCL superconductor. 
%Fig 3

\begin{figure}[b] \begin{center}
	\includegraphics[keepaspectratio=1,width=6 cm, angle = 90]{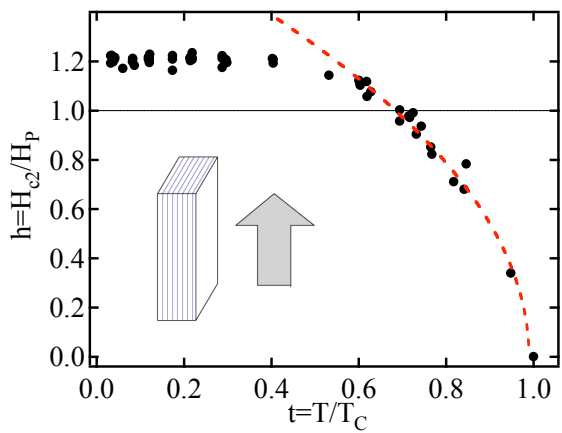} 
	\vspace{-0.2cm}
	\caption{\label{fig:energy}
	The phase diagram, with the applied field parallel to the conducting layers, 
plotted in reduced coordinates ($h=\frac{\hctwo}{\hpBCS}$, $t=\frac{T}{\tc}$).
	The dashed line is a fit to the data where $h\propto \sqrt{1- t}$  
	The solid line indicates $\hpBCS$.  The insert shows the orientation
of the applied field with respect to the conducting layers. } 

\end{center}
\end{figure}

Finally, in Fig.~\ref{fig:energy} we present the phase diagram we measured in reduced
coordinates, where $t=\frac{T}{\tc}$ and  $h=\frac{\hctwo}{\hpBCS}$.  This phase diagram has the essential features of a superconductor in the Pauli limit. The critical field  follows the temperature dependance of the energy gap, starting out with a square root dependance near T$_c$  ($\hctwo\propto(1-t)^{\frac{1}{2}}$) and approaching a constant by t  = 0.4. The low temperature section of this phase diagram is is striking because the low temperature critical fields are saturated at $\unit[2.15]{T}$, which is 20\% above $\hpBCS$, in agreement with our Wilson's ratio calculation.  There is some data in a study using thermal conductivity that supports our phase diagram \cite{ShimojoIshiguro02JPSJ}, and we have recently repeated our experiment with a sample from a different sample grower\cite{quallsNote}, and a new TDO, with results that are within the scatter of the present data. 
%Fig 4

One can estimate $\tau_\text{so}$ using a qualitative comparison between the phase diagram in Fig.~\ref{fig:energy} and Fig.~9 of Ref.~\cite{klemmLB}  and $\tau_\text{so}=\infty$ or if one uses Eq.~(\ref{eq:spinOrbit})  and assumes that all of the enhancement of $\hctwo$ in \ET\ is due to spin-orbit scattering $\tau_\text{so}=\unit[2.0]{ps}$.   In this case it is difficult to know whether the enhancement in $\hctwo$ is due to many body effects, spin orbit scattering or a combination of the two.  Because $\tau_\text{so}$ is at least twice as great as $\tau$, it is more likely that the effects of spin orbit scattering are small to negligible. In the case that spin orbit scattering is negligible, $\Delta_o =  2.0 x 10^{-23}$ J per Cooper pair. 

We would like to acknowledge 
H. Gao, J. Norton, T. Murphy, E. Palm and S. Hannahs for help
doing these experiments, R. Klemm and J. Singleton for useful discussions, 
 L. Rubin for donated equipment from the FBNML, and 
NSF grant \#9805784 and DOE grant ER46214 for support.
	\vspace{-0.7cm}

%Refs
\bibliography{bib}

\begin{thebibliography}{10}
\bibitem{motivation}
The initial motivation for this experiment was two phase diagrams that
  contradicted each other\cite{brooks,shimojo99}.
\end{thebibliography}
\bibliographystyle{PhysRLet}
\end{document}